\documentclass[smallextended,final,numbook]{svjour3}
\journalname{Gen. Rel. Grav.}
 
\usepackage{amssymb}  
\usepackage{amsmath}
\usepackage[usenames,dvipsnames]{xcolor}
\usepackage[normalem]{ulem} 
\usepackage[labelfont=bf,font=small]{caption}
\usepackage{graphicx}
\usepackage{subcaption}

\usepackage{epstopdf}
\usepackage{float}
\usepackage{booktabs} 
\usepackage{microtype}
\usepackage[font=footnotesize]{caption}
\usepackage[colorlinks,breaklinks]{hyperref}


\begin{document} 
 

\title{Simple Numerical Solutions to the Einstein\\ Constraints
  on Various Three-Manifolds}
\titlerunning{Simple Numerical Solutions to the Einstein Constraints}

\author{Fan Zhang \and Lee Lindblom*}
\institute{F. Zhang
  \at {Gravitational Wave and Cosmology Laboratory, Department of Astronomy, Beijing Normal University, Beijing 100875, China \\
              Advanced Institute of Natural Sciences, Beijing Normal University at Zhuhai 519087, China
    \\\email{fnzhang@bnu.edu.cn}}
  \and L. Lindblom (corresponding author)
  \at {Center for Astrophysics and Space Sciences,
    University of California at San Diego, USA
  \\\email{llindblom@ucsd.edu}}}

\date{\today}
 
\maketitle

\vspace{2.0cm}

\begin{abstract}
  Numerical solutions to the Einstein constraint equations are
  constructed on a selection of compact orientable three-dimensional
  manifolds with non-trivial topologies.  A simple constant mean
  curvature solution and a somewhat more complicated non-constant mean
  curvature solution are computed on example manifolds from three of
  the eight Thursten geometrization classes.  The constant mean
  curvature solutions found here are also solutions to the Yamabe
  problem that transforms a geometry into one with constant scalar
  curvature.
\end{abstract}


\section{Introduction} 
\label{s:Introduction} 

Einstein's gravitational field equations are a complicated non-linear
second-order system of partial differential equations for the
components of the spacetime metric.  Like the electromagnetic field
equations, Einstein's equations can be written as a system of
evolution equations plus constraints that must be satisfied at each
instant of time, i.e. on any spacelike surface in the spacetime.
These constraint equations are typically written as systems of
elliptic partial differential equations, which must be solved on an
initial time slice before an evolution can proceed to determine the
full spacetime geometry.  A variety of methods have been developed for
solving these equations on spacetimes of interest to the numerical
relativity community, e.g. for neutron star and black hole binary
systems (see
e.g.~\cite{Cook2000,Pfeiffer2003,Okawa2013,Ossokine2015,Tichy2019}).
This paper focuses on a basic problem that has not received much
attention in the literature to date.  Solutions to the constraints are
explored here on compact orientable three-manifolds having a variety
of different topologies.

Standard numerical relativity codes at this time are not able to solve
problems on manifolds with non-trivial topologies.  Methods have been
developed recently, however, that provide a way to solve partial
differential equations numerically, including the Einstein
constraints, on a wide variety of three-manifolds with different
topologies~\cite{Lindblom2022}.  Those methods are used here to find
simple numerical solutions to the Einstein constraints on four
different manifolds: $S2\times S1$, $G2\times S1$, $L(8,3)$ and
$SFS[S2:(2,1)(2,1)(2,-1)]$.  (The names used for these manifolds are
those used in~\cite{Regina}.) The first two, $S2\times S1$ and
$G2\times S1$, are simple fiber-bundle spaces with $S1$ (the circle)
fibers and base spaces $S2$ (the two-sphere) or $G2$ (the genus two
two-manifold). The $L(8,3)$ manifold is an example of a lens space
obtained from the three-sphere ($S3$) by identifying points related by
a discrete isometry.  The $SFS[S2:(2,1)(2,1)(2,-1)]$ manifold is a
Seifert fibred space constructed from the $S2\times S1$ fiber bundle
by excising neighborhoods of three fibers from this space and twisting
the fibers in these neighborhoods before re-attaching to the $S2$ base
manifold.

Section~\ref{s:EinsteinConstraints} reviews and summarizes the
particular forms of the constraint equations used in this study.
Section~\ref{s:CMCSolutions} describes the simple constant mean
curvature (CMC) solutions to the constraints found here on the example
manifolds described above.  Numerical solutions to this equation are
found using the pseudo-spectral methods implemented in the SpEC code
(developed originally by the Caltech/Cornell numerical relativity
collaboration~\cite{Pfeiffer2003}).  These CMC solutions are also
non-trivial solutions to the Yamabe problem that constructs a constant
scalar curvature geometry on the manifold~\cite{Yamabe1960}.
Section~\ref{s:VMCSolutions} describes the numerically more
challenging and somewhat more complicated non-constant mean curvature
(or variable mean curvature VMC) solutions to the constraints on these
manifolds.  Section~\ref{s:Discussion} summarizes the main results,
and suggests areas where the methods described here might be improved.

%
\section{The Einstein Constraints}
\label{s:EinsteinConstraints}

This section gives a brief introduction to the form of the constraint
equations used in this study.  Consider a spacetime containing a
three-dimensional spacelike surface with future-directed timelike unit
normal $n^\alpha$.\footnote{Greek letters are used for spacetime
indices, e.g. $\alpha$, $\beta$, ..., and Latin letters for spatial
indices on a surface, e.g. $a$, $b$, $c$, ....}  The components of the
Einstein equations,
\begin{eqnarray}
  G_{\alpha\beta}\,n^\alpha n^\beta &=&8\pi T_{\alpha\beta}\,n^\alpha n^\beta,
  \label{e:Einstein1}\\
  G_{a\beta}\,n^\beta&=& 8\pi T_{a\beta}\,n^\beta,
  \label{e:Einstein2}
\end{eqnarray}
play the role of initial value constraints on this surface. When
re-written in terms of the spatial metric $g_{ab}$ and extrinsic
curvature $K_{ab}$ of this surface, these equations have the form,
\begin{eqnarray}
  G_{\alpha\beta}\,n^\alpha n^\beta &=&\textstyle\frac{1}{2}\left(R - K_{ab}K^{ab} + K^2\right)
  =  8\pi T_{\alpha\beta}\,n^\alpha n^\beta,
  \label{e:HamiltonianConstraint}\\
  G_{a\beta}\,n^\beta&=&\nabla^bK_{ba} - \nabla_a K = 8\pi T_{a\beta}n^\beta,
  \label{e:MomentumConstraint}
\end{eqnarray}  
where $R$ is the scalar curvature associated with the metric $g_{ab}$,
$\nabla_a$ is the $g_{ab}$ metric-compatible covariant
derivative, and $K=g^{ab}K_{ab}$ on this surface.

The most general and most widely used method of solving these constraints
re-expresses $g_{ab}$ and $K_{ab}$ in terms of ``conformal'' fields
$\phi$, $\tilde g_{ab}$, $\tilde\tau$, $\tilde\sigma_{ab}$ and $\tilde W_a$
(for a review see \cite{BartnikIsenberg2004}):
\begin{eqnarray}
  g_{ab} &=& \phi^4 \tilde g_{ab},
  \label{e:ConformalMetric}\\
  K_{ab}&=&\phi^{-2}(\tilde\sigma_{ab}+\widetilde{LW}_{ab})
  +\textstyle\frac{1}{3}\phi^4 \tilde g_{ab}\tilde\tau,
  \label{e:ConformalExtrinsicCurvature}
\end{eqnarray}
where $\phi>0$ is the conformal factor, $\tilde g_{ab}$ is a positive
definite metric, $\tilde\sigma_{ab}$ is trace-free and divergence-free
(with respect to the $\tilde g_{ab}$ metric-compatible covariant
derivative $\tilde\nabla_a$), and $\tilde \tau=K$.  The tensor
$\widetilde{LW}_{ab}$ is defined as the shear of $\tilde W_a$:
\begin{equation}
  \widetilde{LW}_{ab}=\tilde\nabla_a\tilde W_b+\tilde\nabla_b\tilde W_a
  - \textstyle\frac{2}{3}\tilde
  g_{ab}\tilde\nabla_c\tilde W^c.
\end{equation}

The constraints, Eqs.~(\ref{e:HamiltonianConstraint}) and
(\ref{e:MomentumConstraint}), can be re-written as a system of
equations for $\phi$ and $\tilde W_a$ by using the following identities that
relate the covariant derivative $\nabla_a$ and $\tilde\nabla_a$ (the
covariant derivative compatible with the conformal metric $\tilde
g_{ab}$):
\begin{eqnarray}
  \nabla^a\rho_{ab}&=&\phi^{-6}\,\tilde\nabla^a(\phi^2\rho_{ab}),
  \label{e:ConformalDivergences}\\
  R &=& \phi^{-4}\, \tilde R - 8 \phi^{-5}\, \tilde \nabla^a\tilde\nabla_a\phi,
  \label{e:ConformalRicciScalar}
\end{eqnarray}
where $\rho_{ab}$ is any trace-free symmetric tensor field, and
$\tilde R$ is the scalar curvature associated with $\tilde g_{ab}$.
Using these identities Eqs.~(\ref{e:HamiltonianConstraint}) and
(\ref{e:MomentumConstraint}) can be written as,
\begin{eqnarray}
&&  \tilde\nabla^a\tilde\nabla_a\phi = \textstyle\frac{1}{8}\phi\, \tilde R
  +\textstyle\frac{1}{12}\phi^5\, \tilde\tau^2 
  -\textstyle\frac{1}{8}\phi^{-7}
  (\tilde\sigma_{ab}+\widetilde{LW}_{ab})
  (\tilde\sigma^{ab}+\widetilde{LW}^{ab})- 2\pi \phi^5 T_{\perp\perp},
  \nonumber\\
    \label{e:LichnerowitzEq}\\
&&  \tilde\nabla^b(\widetilde{LW}_{ba})=
  \textstyle\frac{2}{3}\phi^6\,\tilde\nabla_a\tilde\tau
  + 8\pi\phi^6 T_{a\perp},
  \label{e:MomentumConstraint2}
\end{eqnarray}
where $T_{\perp\perp}=T_{\alpha\beta}\,n^\alpha n^\beta$ and
$T_{a\perp}=T_{a\beta}\,n^\beta$.  The stress-energy components
$T_{\perp\perp}$ and $T_{a\perp}$ are determined by the physical
properties of the matter in the spacetime, while the conformal fields
$\tilde g_{ab}$, $\tilde \sigma_{ab}$, $\tilde \tau$ can be chosen
freely.  Once these stress-energy and conformal fields are fixed,
Eqs.~(\ref{e:LichnerowitzEq}) and (\ref{e:MomentumConstraint2}) become
a second-order system of elliptic equations for $\phi$ and $\tilde
W_a$.

Differentiable structures were constructed numerically
in~\cite{Lindblom2022} for a collection of forty different
three-manifolds having representative topologies from five of the
eight Thurston geometrization classes~\cite{Thurston1997,Scott1983}.
The goal here is to construct simple solutions to
Eqs.~(\ref{e:LichnerowitzEq}) and (\ref{e:MomentumConstraint2})
numerically on a selection of those manifolds.  The procedure
introduced in~\cite{Lindblom2022} produces a $C^1$ reference metric
$\tilde g_{ab}$ on these manifolds.  Those reference metrics are used
to construct Jacobians and a covariant derivative that define what it
means for tensor fields to be continuous and differentiable across the
boundaries between coordinate patches.  These reference metrics are
also used here as the conformal metric that appears in
Eqs.~(\ref{e:LichnerowitzEq}) and (\ref{e:MomentumConstraint2}).

The symmetric trace-free divergence-free tensor $\tilde \sigma_{ab}$
is often associated with gravitational-wave degrees of freedom.  The
differentiable structures constructed in~\cite{Lindblom2022} for
these example manifolds provide no structure from which a suitable
$\tilde \sigma_{ab}$ could easily be constructed.  Therefore for
simplicity the solutions constructed here set $\tilde \sigma_{ab}=0$.

Another common simplification used in the solution to the Einstein
constraints is to set $\tilde\nabla_a\tilde\tau=0$.  In this case the
topologies of the manifolds on which vacuum solutions exist,
i.e. those with $T_{\perp\perp}=T_{\perp\kern 0.15em a}=0$, are known
to be limited~\cite{Isenberg1995}.  To avoid this restriction, a
very simple form of matter is introduced to allow solutions to exist
for all the cases considered here.  In particular a cosmological
constant $\Lambda$ is included, whose stress energy tensor is given
by,
\begin{equation}
  T_{\alpha\beta} = -\frac{\Lambda}{8\pi}\psi_{\alpha\beta},
\end{equation}
where $\psi_{\alpha\beta}$ is the full spacetime-metric.  The
components $T_{\perp\perp}$ and $T_{\perp\kern 0.15em a}$ that enter
the constraints in this case, are given by,
\begin{eqnarray}
  T_{\perp\perp}&=& \frac{\Lambda}{8\pi},\\
  T_{\perp\kern 0.15em a} & = & 0.
\end{eqnarray}
These assumptions simplify the structures of
Eqs.~(\ref{e:LichnerowitzEq}) and (\ref{e:MomentumConstraint2}):
\begin{eqnarray}
  \tilde\nabla^a\tilde\nabla_a\phi &=& \textstyle\frac{1}{8}\phi\, \tilde R
  +\textstyle\frac{1}{12}\phi^5\, (\tilde\tau^2-3\Lambda)
  -\frac{1}{8}\phi^{-7}\widetilde{LW}_{ab}\,\widetilde{LW}^{ab}
,
  \label{e:SimpleConstraintS}\\
  \tilde\nabla^b(\widetilde{LW}_{ba})&=&
  \textstyle\frac{2}{3}\phi^6\,\tilde\nabla_a\tilde\tau.
  \label{e:SimpleConstraintV}
\end{eqnarray} 
Two classes of simple solutions to these equations are constructed
numerically in the following sections: those with
$\tilde\nabla_a\tilde\tau=0$ (the constant mean curvature solutions)
in Sec.~\ref{s:CMCSolutions}, and those with
$\tilde\nabla_a\tilde\tau\neq 0$ (the variable mean curvature
solutions) in Sec.~\ref{s:VMCSolutions}.

An important way to measure how well the numerical solutions
successfully solve Eqs.~(\ref{e:SimpleConstraintS}) and
(\ref{e:SimpleConstraintV}) is to evaluate how well they satisfy the
original Einstein constraints Eqs.~(\ref{e:HamiltonianConstraint}) and
(\ref{e:MomentumConstraint}).  To do that the physical metric $g_{ab}$
and extrinsic curvature $K_{ab}$ are re-constructed from the
numerically determined $\phi$ and $\tilde W_a$ using
Eqs.~(\ref{e:ConformalMetric}) and
(\ref{e:ConformalExtrinsicCurvature}).  The scalar curvature $R$
associated with $g_{ab}$ is then determined numerically, which allows
the original forms of the Hamiltonian $\mathcal{H}$ and momentum
$\mathcal{M}_{\,a}$ constraints, Eqs.~(\ref{e:HamiltonianConstraint})
and (\ref{e:MomentumConstraint}), to be evaluated,
\begin{eqnarray}
  \mathcal{H} & = & R - K_{ab}K^{ab} + K^2 - 16\pi\, T_{\perp\perp},
  \label{e:HamiltonianConstraint2}\\
  \mathcal{M}_{\,a} & = & \nabla^bK_{ba} - \nabla_a K - 8\pi\, T_{\perp\kern 0.15em a}.
  \label{e:MomentumConstraint3}
\end{eqnarray}
The accuracy of the resulting $g_{ab}$ and $K_{ab}$ can then be
measured using the following constraint norm,
\begin{eqnarray}
  \mathcal{C}^2 = \mathcal{V}^{-1}\int \left(\mathcal{H}^2
  + g^{ab}\mathcal{M}_{\,a}\mathcal{M}_{\,b}\right)\!\sqrt{\det g}\, d^3x,
  \label{e:ConstraintNorm}
\end{eqnarray}
where $\mathcal{V}$ is the proper volume of the manifold,
\begin{equation}
  \mathcal{V} = \int \sqrt{\det g}\, d^3x.
  \label{e:PhysicalVolumeV}
\end{equation}
This norm, $\mathcal{C}$, vanishes for an exact solution to the
Einstein constraints, so a non-zero value is a useful measure of the
accuracy of a numerical solution.


\section{Simple Constant Mean Curvature (CMC) Solutions}
\label{s:CMCSolutions}

This section defines a simple one parameter family of constant mean
curvature (CMC) solutions to the Einstein constraints, and reports the
results of numerical evaluations of these solutions on a selection of
three-dimensional manifolds with different topologies.

In the constant mean curvature case, $\tilde\nabla_a\tilde\tau=0$, the
Einstein constraints Eqs.~(\ref{e:SimpleConstraintS}) and
(\ref{e:SimpleConstraintV}) simplify considerably.  In particular
Eq.~(\ref{e:SimpleConstraintV}) becomes a homogeneous elliptic
equation for $\tilde W_a$, $\tilde\nabla^b (\widetilde{LW}_{ab})=0$,
whose simplest (and in most cases unique\footnote{The $W_a=0$ solution
is unique up to the addition of a conformal Killing field, and none
exist for most geometries.}) solution is $\tilde W_a=0$.  This in turn
reduces Eq.~(\ref{e:SimpleConstraintS}) to the following,
\begin{eqnarray}
  \tilde\nabla^a\tilde\nabla_a\phi &=& \textstyle\frac{1}{8}\phi\, \tilde R
  +\textstyle\frac{1}{12}\phi^5\, (\tilde\tau^2-3\Lambda).
  \label{e:CMCConstraintS}
\end{eqnarray}
The integral of the left side of Eq.~(\ref{e:CMCConstraintS}) vanishes
on any compact manifold. Therefore the constants $\tilde\tau$ and
$\Lambda$ must be chosen in a way that makes it possible for the
integral of the right side of this equation to vanish as well.
Convenient choices for these constants would produce
  solutions to Eq.~(\ref{e:CMCConstraintS}) with $\phi\approx 1$.
  Such choices can be identified by setting $\phi=1$ in the
  expression on the right side of Eq.~(\ref{e:CMCConstraintS}) and
  integrating over the manifold.  Setting this integral to zero
  results in the values, 
\begin{equation}
  \tilde\tau^2 - 3\Lambda = - \textstyle\frac{3}{2}\langle \tilde R\, \rangle,
  \label{e:CMCParameterChoices}
\end{equation}
where $\langle \tilde R\, \rangle$ is the average value of the
conformal scalar curvature $\tilde R$,
\begin{equation}
  \langle \tilde R\, \rangle
  = \frac{\int \sqrt{\det\tilde g}\,\tilde R\, d^{\,3}x}
  {\int \sqrt{\det\tilde g}\, d^{\,3}x}.
  \label{e:AvarageRtildeDef}
\end{equation}
This choice transforms Eq.~(\ref{e:CMCConstraintS}) into the form
\begin{eqnarray}
  \tilde\nabla^a\tilde\nabla_a\phi &=& \textstyle\frac{1}{8}\phi\,
  \left(\tilde R - \phi^4 \langle \tilde R\,\rangle\right).
  \label{e:CMCConstraintSAlt}
\end{eqnarray}
This equation has the exact solution $\phi=1$ in the constant scalar
curvature case $\tilde R =\langle \tilde R\,\rangle$, and admits
solutions in all the CMC cases studied here. The integral of the right
side of Eq.~(\ref{e:CMCConstraintS}) must vanish for any solution
$\phi$.  If $\phi>0$ and $\tilde R>0$ this integral can vanish only if
$\tilde\tau^2 - 3\Lambda<0$.  Thus no $\phi>0$ solution can exist to
Eq.~(\ref{e:CMCConstraintS}) when $\tilde R>0$ unless the cosmological
constant satisfies the inequality, $\Lambda > \tfrac{1}{3}\tilde\tau^2
\geq 0$.

Once a conformal metric $\tilde g_{ab}$ is chosen,
Eq.~(\ref{e:CMCConstraintSAlt}) becomes a second-order elliptic
differential equation that can be solved using a variety of standard
numerical methods.  The conformal metrics used for the examples in
this study are the reference metrics constructed
in~\cite{Lindblom2022} for building differentiable structures on these
manifolds.  These positive-definite metrics are smooth within each
cubic coordinate chart, and are continuous and differentiable in the
appropriate senses across the interfaces between charts.

Table~\ref{t:TableI} lists the compact orientable manifolds selected
for this study, $S2\times S1$, $G2\times S1$, $L(8,3)$ and
$SFS[S2:(2,1)(2,1)(2,-1)]$, which are described in physical terms
briefly in Sec.~\ref{s:Introduction}. These three-manifolds belong to
three different Thursten geometrization classes: $L(8,3)$ and
$SFS[S2:(2,1)(2,1)(2,-1)]$ belong to the $S^3$ class, $G2\times S1$
belongs to the $H^2\times S^1$ class, and $S2\times S1$ is the
defining member of the $S^2\times S^1$ class.  This table also lists
$\langle \tilde R\,\rangle$ defined in Eq.~(\ref{e:AvarageRtildeDef})
and the physical volumes $\mathcal{V}(\mathrm{CMC})$ defined in
Eq.~(\ref{e:PhysicalVolumeV}) for the CMC geometries constructed in
this study on each of these manifolds.  These volumes measure the
physical ``sizes'' of the manifolds in the length-scale units of our
code, and can therefore be used to calibrate the sizes of the
curvatures of the geometries.
\begin{table}[!hbt]
  \caption{Compact orientable manifolds included in this study. Also
    listed are the average scalar curvature $\langle \tilde R\rangle$
    defined in Eq.~(\ref{e:AvarageRtildeDef}), and the physical
    volumes $\mathcal{V(\mathrm{CMC})}$ defined in
    Eq.~(\ref{e:PhysicalVolumeV}) for the CMC geometries constructed
    on each manifold.
    \label{t:TableI} }
  \begin{center}
  \begin{tabular}{lcc}
    Manifold & $\langle \tilde R\, \rangle$
    & $\mathcal{V}(\mathrm{CMC})$
 \\
    \hline
    $G2\times S1$              & -2.97 & \,9.68  \\
    $L(8,3)$                   &  2.66 & \,8.23  \\
    $S2\times S1$              &  2.69 & \,9.42  \\
    $SFS[S2:(2,1)(2,1)(2,-1)]$ &  2.66 & \,8.23  \\
  \end{tabular}
  \end{center}
\end{table}

Numerical solutions of Eq.~(\ref{e:CMCConstraintSAlt}) are constructed
in this study using multicube representations of these manifolds, as
described in \cite{Lindblom2022,Lindblom2013}.  A multicube
representation is a collection of non-overlapping cubes in
$\mathbb{R}^3$ together with maps that specify how the faces are
identified with the faces of neighboring cubes.  These multicube
regions serve as the coordinate charts used to represent tensor fields
on these manifolds. Complete descriptions of the multicube structures
used for each of the manifolds included in this study are included in
Appendix A.  Figure~\ref{f:ReferenceMetic} illustrates the multicube
structure used to represent the $G2\times S1$ manifold, with surface
colors representing $\sqrt{\det \tilde g}$ and $\tilde R$ for the
reference metric used here.  Blue colors in these figures represent
small values of these scalars, and red colors represent large values.
The scalar curvatures $\tilde R$ for the reference metrics used in
this study are not constant, as illustrated in
Fig.~\ref{f:ScalarRtilde}.  Therefore the constraint
Eq.~(\ref{e:CMCConstraintSAlt}) is not trivial even in the simple CMC
case studied here.
\begin{figure}[!h]
  \begin{subfigure}{0.5\textwidth}
      \hspace{-0.4cm}
    \includegraphics[width=0.98\textwidth]{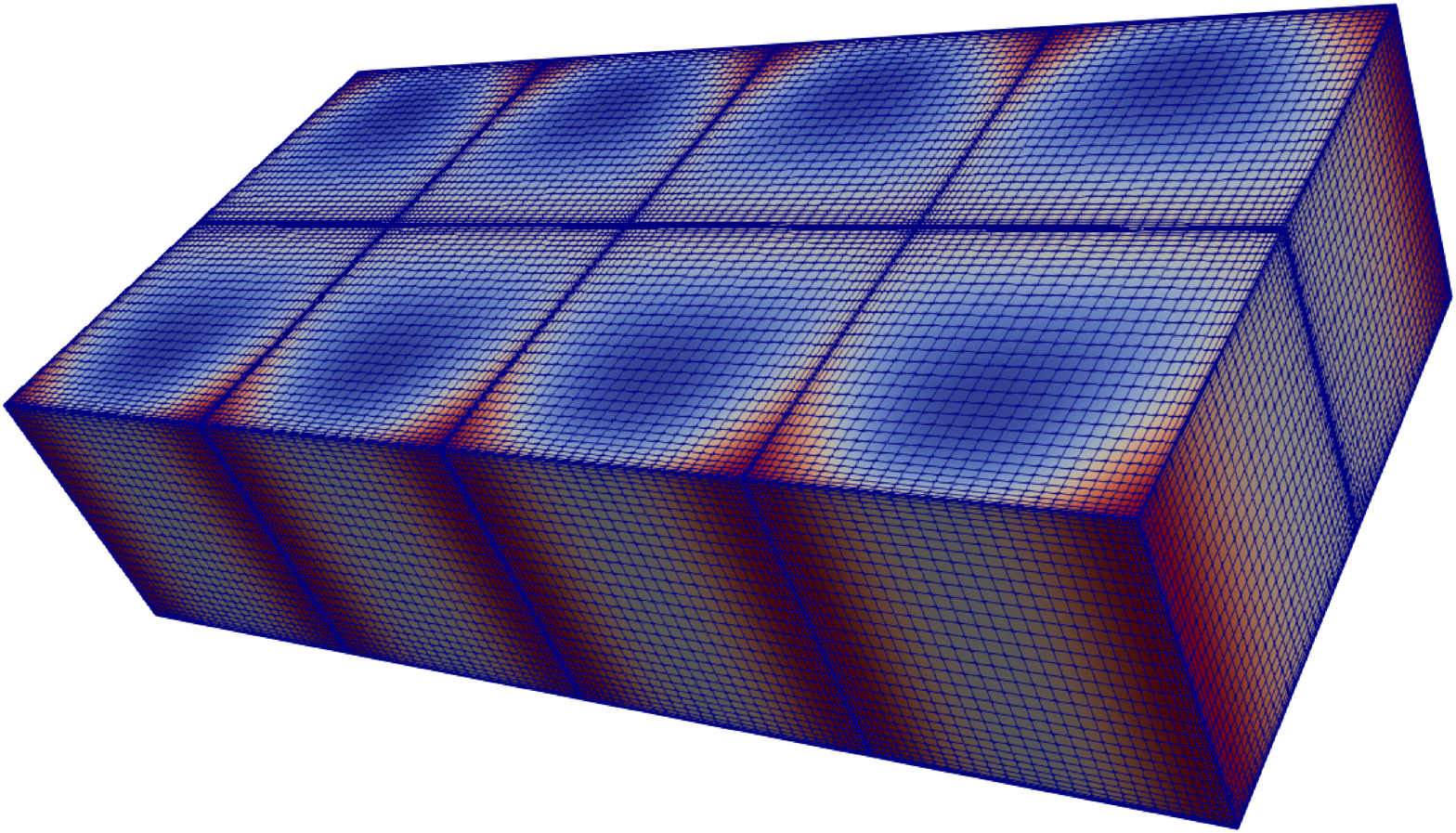}
    \caption{\label{f:Detgtilde}Square root of determinant of $\tilde g_{ab}$.}
  \end{subfigure}
  \hspace{-0.4cm}
  \begin{subfigure}{0.5\textwidth}
  \hspace{-0.4cm}
    \includegraphics[width=0.98\textwidth]{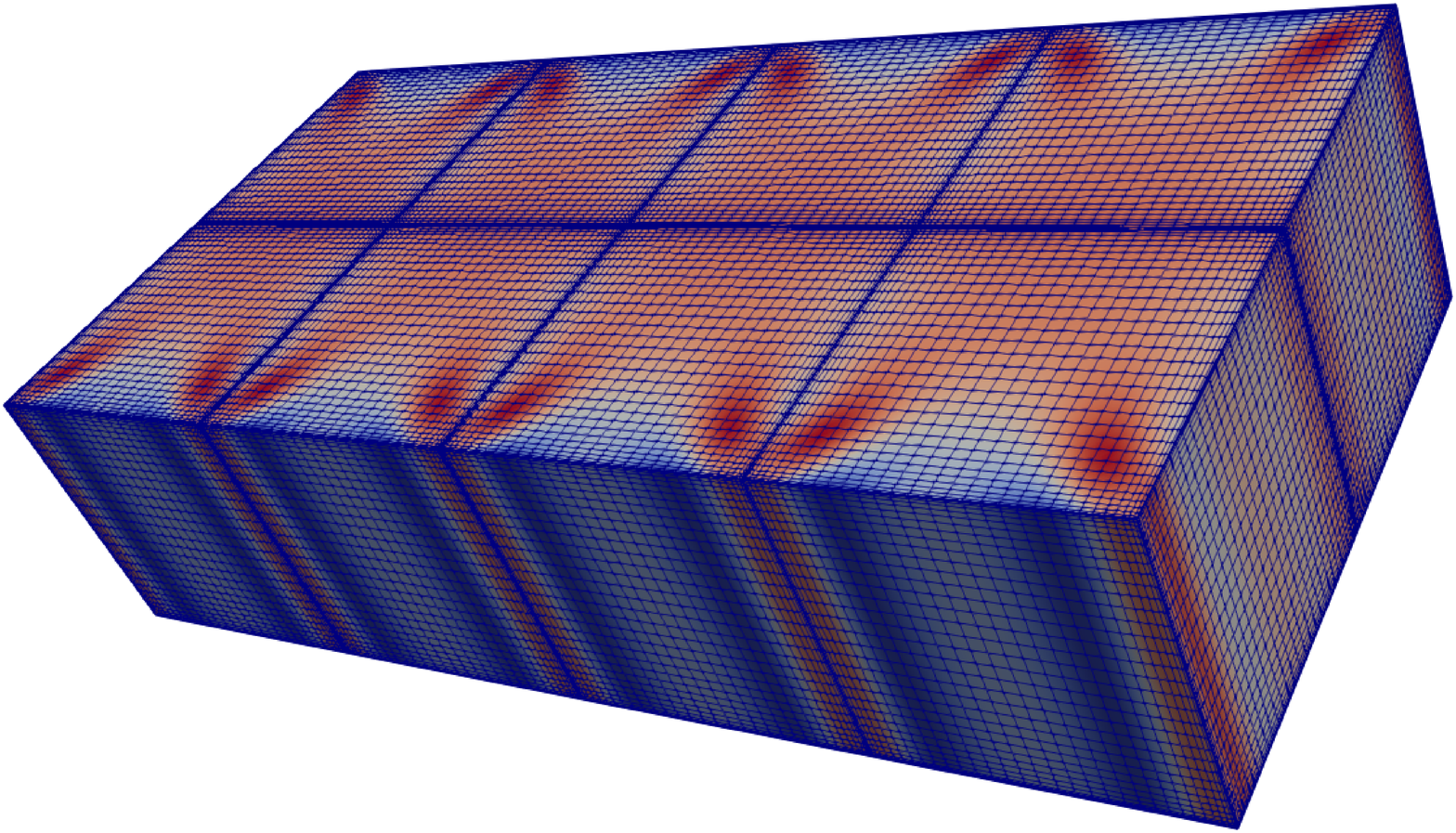}
    \caption{\label{f:ScalarRtilde}Scalar curvature $\tilde R$.}
  \end{subfigure}
  \caption{Views of the multicube structure used to represent the
    $G2\times S1$ manifold, along with the surface values of the
    determinant of the reference metric, $\sqrt{\det \tilde g}$, and the
    scalar curvature $\tilde R$.
    \label{f:ReferenceMetic}}
\end{figure}

For this study the differential Eq.~(\ref{e:CMCConstraintSAlt}) has
been solved numerically using the pseudo-spectral methods implemented
in the SpEC numerical relativity code~\cite{Pfeiffer2003}.  Functions
are represented by their values on a grid defined by the locations of
the Gauss-Lobatto collocation points.  Representing functions in this
way provides a numerically efficient way to transform back and forth
between the grid representation of functions, and their representation
as Chebyshev polynomial expansions.  Derivatives are evaluated
numerically using the exact analytic expressions for the derivatives
of those Chebyshev expansions.  The elliptic differential operator in
Eq.~(\ref{e:CMCConstraintSAlt}) becomes in effect a linear matrix that
operates on the vector of grid values of $\phi$.  Boundary conditions
are included in this matrix operator by replacing the equation for the
elliptic operator by equations that enforce the continuity of $\phi$
and its gradient $\nabla\phi$ on the grid points along the interface
boundaries between the multicube coordinate charts.  Details about how
the SpEC code implements these boundary conditions can be found in
\cite{Pfeiffer2003} and more specifically for multicube manifolds in
Sec.~5 of \cite{Lindblom2013}.  The non-linear
Eq.~(\ref{e:CMCConstraintSAlt}) is solved in effect by minimizing the
discrete version of the residual $\mathcal{E}$ defined by
\begin{equation}
  \mathcal{E}=\tilde\nabla^a\tilde\nabla_a\phi
  -\textstyle\frac{1}{8}\phi\, \left(\tilde R
  - \phi^4 \langle \tilde R\,\rangle\right).
  \label{e:CMCResidualDef}
\end{equation}

The SpEC code minimizes these residuals by accessing the ksp linear
solver and the snes non-linear solver from the PETSC software
library~\cite{petsc-user-ref}.  These solves are done iteratively,
starting with the initial guess $\phi=1$ for the lowest spatial
resolution.  Once the solver finds a solution that satisfactorily
minimizes $\mathcal{E}$ for one resolution, that solution is
interpolated onto the next higher resolution grid as its initial
guess.  This procedure is repeated through a series of increasing
numerical resolutions.  Solving the equation in this way mimics the
advantages of a multi-grid solver by allowing the long length-scale
features of the solution (which take the longest to converge
numerically) to be determined in the faster low-resolution solves.
The CMC solutions for this study have been computed on a sequence of
grids with $N=\{16,20,24,28,32,35\}$ collocation points in each
spatial direction in each multicube region.  The SpEC code
parallelizes these computations (up to a point) by allowing each
multicube region to be run on a separate processor.  These numerical
computations take a very long time, and this has limited our ability
to consider additional example manifolds or to explore them with
higher numerical resolutions.

\begin{figure}[!h] 
  \centering
    \includegraphics[width=0.48\textwidth]{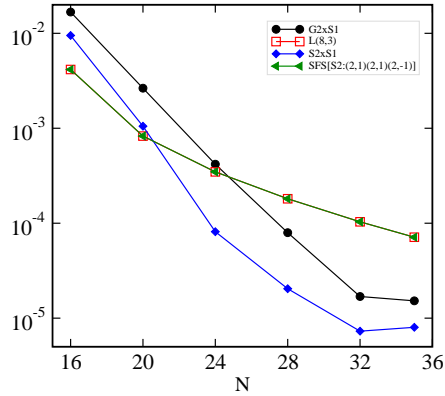}
  \caption{\label{f:CMCEinsteinConstraints}Norm of the Einstein
    constraints, $\mathcal{C}$, as functions of the numerical
    resolution $N$ for the numerical CMC initial data solutions.}
\end{figure}

The constraint norm $\mathcal{C}$ defined in
Eq.~(\ref{e:ConstraintNorm}) vanishes for any exact solution to the
constraint equations and is therefore an important and useful measure
of the accuracy of the numerical solutions.  This constraint norm has
a particularly simple form for these simple CMC solutions.  The
momentum constraint from Eq.~(\ref{e:MomentumConstraint3}) is
satisfied identically in this case, $\mathcal{M}_{\,a}=0$, since
$\tilde \sigma_{ab}=\tilde W_a=0$.  Thus $\mathcal{C}$ depends only on
the Hamiltonian constraint $\mathcal{H}$ defined in
Eq.~(\ref{e:HamiltonianConstraint2}).  For the CMC case $\mathcal{H}$
is given by
\begin{equation}
  \mathcal{H}=R +\textstyle\frac{2}{3}\left(\tilde\tau^2 - 3\Lambda\right) = R
  - \langle\tilde R\rangle.
\end{equation}
Consequently the constraint norm $\mathcal{C}$ becomes
\begin{equation}
  \mathcal{C}^2 = \mathcal{V}^{-1}\int \left(R - \langle\tilde R\,
  \rangle\right)^2 \!\sqrt{\det g}\, d^3x.
\end{equation}
The vanishing of $\mathcal{C}$ implies that the scalar curvature $R$
is constant, $R=\langle\tilde R\,\rangle$, for these simple CMC
solutions.  Thus the conformal factor $\phi$ is the solution to the
Yamabe problem that transforms $\tilde g_{ab}$ into the constant
scalar curvature metric $g_{ab}$~\cite{Yamabe1960}.
Figure~\ref{f:CMCEinsteinConstraints} illustrates the values of the
constraint $\mathcal{C}$ as a function of the spatial resolution $N$
(the number of grid points in each direction of each multicube region)
for each of the manifolds studied here.  These results show that our
numerical methods (generally) converge with increasing values of the
spatial resolution $N$, and produce reasonably accurate solutions to
the constraint equations. The values of $\mathcal{C}$ for the $N=35$
resolutions of the $G2\times S1$ and $S2\times S1$ manifolds are
larger than expected. These numerical solutions are very time
consuming for the higher resolution cases, and it is possible that the
final results reported here could have been improved somewhat with
more computer time or perhaps by setting somewhat different parameters
in the PETSC solvers.

\begin{figure}[!h]
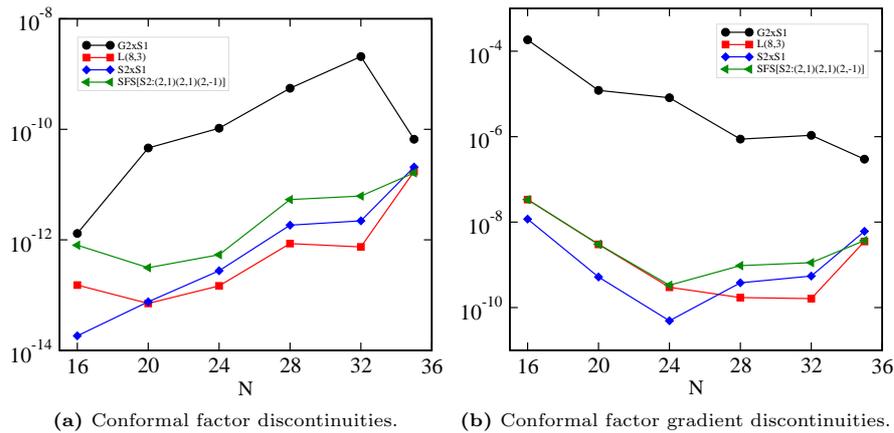

  \begin{subfigure}{0.49\textwidth}
    \centering
  \includegraphics[width=0.98\textwidth]{Fig3.2a.eps}
  \caption{\label{f:CMCConformalFactor}Conformal factor discontinuities.}
\end{subfigure}
  \begin{subfigure}{0.49\textwidth} 
    \centering
    \includegraphics[width=0.98\textwidth]{Fig3.2b.eps}
    \caption{\label{f:CMCConformalFactorGradient}Conformal factor gradient
      discontinuities.}
  \end{subfigure}
  \caption{Norms of the interface discontinuities in the conformal
    factor and its gradient as functions of the spatial resolution $N$
    for the CMC solutions.
    \label{f:CMCDiscontinuities}}
\end{figure}

Solutions to the CMC Einstein constraint
Eq.~(\ref{e:CMCConstraintSAlt}) should be smooth across the interface
boundaries between multicube coordinate patches.  Therefore the
continuity of the resulting solutions and their derivatives across
those interface boundaries is another basic measure of how well these
numerical solutions successfully solve the constraint equations
globally.  The $L_2$ norms of the differences between these boundary
values of the conformal factor $\phi$ are computed by taking the
square root of the squares of the differences averaged over all the
boundary grid points.  These norms are shown in
Fig.~\ref{f:CMCDiscontinuities} for each numerical resolution $N$ for
each of the manifolds studied here.  The results show that the
numerical CMC solutions have boundary continuity errors that are
orders of magnitude smaller than the Einstein constraint errors for
these solutions shown in Fig.~\ref{f:CMCEinsteinConstraints}.  These
discontinuity errors therefore do not contribute significantly to
the Einstein constraint errors for these solutions.
%

\section{Simple Variable Mean Curvature (VMC) Solutions}
\label{s:VMCSolutions}

This section defines a simple one parameter family of variable mean
curvature (VMC) solutions to the Einstein constraints, and reports the
results of numerical evaluations of these solutions on a selection of
three-dimensional manifolds with different topologies.

The constraint equations in the simple VMC case studied here are given
by,
\begin{eqnarray}
  \tilde\nabla^a\tilde\nabla_a\phi &=& \textstyle\frac{1}{8}\phi\, \tilde R
  +\textstyle\frac{1}{12}\phi^5\, (\tilde\tau^2-3\Lambda)
  -\frac{1}{8}\phi^{-7}\widetilde{LW}_{ab}\,\widetilde{LW}^{ab}
,
  \label{e:SimpleConstraintS2}\\
  \tilde\nabla^b(\widetilde{LW}_{ba})&=&
  \textstyle\frac{2}{3}\phi^6\,\tilde\nabla_a\tilde\tau.
  \label{e:SimpleConstraintV2}
\end{eqnarray} 
These become a second-order system of elliptic equations for $\phi$
and $\tilde W_a$ once the conformal fields $\tilde g_{ab}$,
$\tilde\sigma_{ab}$, $\tilde\tau$ and the cosmological constant
$\Lambda$ are chosen.  Unlike the CMC case, these equations are
coupled so they must be solved as a single large system rather than
individually one after the other.

The conformal fields $\tilde g_{ab}$ and $\tilde \sigma_{ab}$ for
these simple VMC solutions are chosen to be the same as those used for
the CMC solutions described in Sec.~\ref{s:CMCSolutions}. The
conformal metric $\tilde g_{ab}$ is identified with the reference
metric constructed using the methods describe in~\cite{Lindblom2022}
for that manifold.  The transverse trace-free part of the conformal
extrinsic curvature, $\tilde\sigma_{ab}$ is set to zero.  Given these
choices, the only remaining freedoms are the choices of a suitable
non-constant $\tilde \tau$ and the cosmological constant $\Lambda$.

\begin{figure}[!ht]
  \centering
  \includegraphics[width=0.48\textwidth]{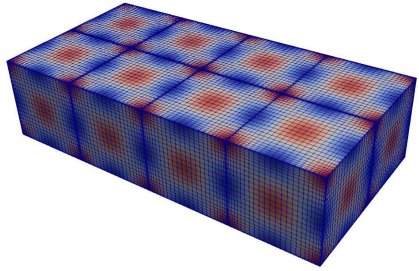}
  \caption{\label{f:TauTilde} Surface values of $\tilde\tau$ for the
    simple VMC solution on the $G2\times S1$ manifold.  The scale of
    the spatial variations in $\tilde\tau$ is set by the variance
    parameter $\mu$.  In the example shown here $\mu=0.1$ which has
    $\max\tilde\tau/\min\tilde\tau\approx 1.8$.}
\end{figure}

The only requirements on $\tilde \tau$ are that it must be continuous
and differentiable, even across the interfaces between multicube
regions, and sufficiently slowly varying to be easily resolved by the
numerical code.  One possibility is to set
\begin{equation}
  \tilde \tau = A\,\Bigl(1 + B\, h(s^x)\, h(s^y)\, h(s^z)\Bigr),
  \label{e:TildeTauDef}
\end{equation}
where $A$ and $B$ are constants, $s^x$, $s^y$ and $s^z$ are re-scaled
local coordinates in each multicube coordinate chart with ranges $-1
\leq s^x, s^y, s^z \leq 1$, and $h(s)$ is defined by
\begin{equation}
  h(s) = \textstyle\frac{8}{15}-(1-s^2)^2.
\end{equation}
This $h(s)$ has the value $h(\pm 1)=\frac{8}{15}$ and derivative
$\frac{dh(\pm 1)}{ds}=0$ on each of the boundaries of the coordinate
patch where $s^2=1$.  (The $\frac{8}{15}$ constant was chosen to make
the integral of $h(s)$ vanish.) Therefore $\tilde\tau$ defined in
Eq.~(\ref{e:TildeTauDef}) is continuous and differentiable in the
appropriate sense for any values of the global constants $A$ and $B$.
The spatial average of $\tilde\tau$ is $\langle\tilde\tau\rangle=A$,
so a natural choice for $A$ is $A^2 = |\langle\tilde R\rangle|$, which
makes the scale of the extrinsic curvature comparable to the scale of
the scalar curvature $\tilde R$.  The spatial variation in
$\tilde\tau$ is determined by $B$. The variance $\mu$ is defined as
the rms average spatial variation in $\tilde\tau$, and is related to
$B$ by $B^2=\mu^2\left(\frac{525}{64}\right)^3$.  Using these choices
for $A$ and $B$ produces the $\tilde\tau$ used here for the simple VMC
solutions:
\begin{equation}
  \tilde\tau = \bigl|\langle\tilde R\rangle\bigr|^{1/2}\left[
    1+\mu\left(\textstyle\frac{525}{64}\right)^{3/2}h(s^x)h(s^y)h(s^z)\right].
  \label{e:TildeTauSimpleVMC}
\end{equation} 
Figure~\ref{f:TauTilde} illustrates the surface values of this
$\tilde\tau$ on the multicube structure used to represent the
$G2\times S1$ manifold in this study.  The variance parameter
$\mu=0.1$ used for the example in this figure results in spatial
variations of $\tilde \tau$ with $\max\tilde\tau/\min\tilde\tau\approx
1.8$.

The last choice needed to fix these simple VMC solutions is the value
of the cosmological constant $\Lambda$.  If a solution to
Eq.~(\ref{e:SimpleConstraintS2}) exists, the integral of its right
side must vanish.  The idea is to choose $\Lambda$ that makes it
possible to have solutions with $\phi\approx 1$.  In this case the
spatial average of the terms on the right side of
Eq.~(\ref{e:SimpleConstraintS2}) must satisfy,
\begin{equation}
  0\approx \textstyle\frac{1}{8}\langle \tilde R\,\rangle
  -\textstyle\frac{1}{8}\langle \widetilde{LW}_{ab}\,\widetilde{LW}^{ab}\rangle
  +\textstyle\frac{1}{12}\langle\tilde \tau^2\rangle -\frac{1}{4}\Lambda.
  \label{e:ApproxLambdaDef}
\end{equation}
From Eq.~(\ref{e:SimpleConstraintV2}) it follows that the spatial
variations in $\widetilde{LW}_{ab}$ should be comparable in size to
the spatial variations in $\tilde\tau$, i.e. $\langle
\widetilde{LW}_{ab}\widetilde{LW}^{ab}\rangle
\approx\frac{4}{9}\mu^2\langle\tilde\tau\rangle^2$.  The quantity
$\langle\tilde\tau^2\rangle$ that appears in
Eq.~\ref{e:ApproxLambdaDef} is also determined by the spatial
variation in $\tilde\tau$: $\langle\tilde\tau^2\rangle
=(1+\mu^2)\langle\tilde\tau\rangle^2$. Thus a suitable choice for
$\Lambda$ should be
\begin{eqnarray}
  \Lambda   &=&  \textstyle\frac{1}{2}\langle\tilde R\rangle
  +\textstyle\frac{1}{9}(3+\mu^2)|\langle\tilde R\rangle|.
  \label{e:LambdaSimpleVMC}
\end{eqnarray}

The simple VMC solutions described above were constructed in this
study by solving Eqs.~(\ref{e:SimpleConstraintS2}) and
(\ref{e:SimpleConstraintV2}) numerically.  These solutions were
obtained for each of the manifolds listed in Table~\ref{t:TableII}
using the numerical methods described in Sec.~\ref{s:CMCSolutions} for
the CMC case.  The expression used for $\tilde\tau$ in these solutions
is given in Eq.~(\ref{e:TildeTauSimpleVMC}).  The variance parameter
in this expression is set to $\mu=0.1$ for the solutions on the
$G2\times S1$ and the $S2\times S1$ manifolds, and $\mu=0.01$ for the
$L(8,3)$ and $SFS[S2:(2,1)(2,1)(2,-1)]$ manifolds to speed up
convergence in those cases.  The cosmological constant $\Lambda$ used 
for these simple VMC solutions is given in
Eq.~(\ref{e:LambdaSimpleVMC}). 
\begin{table}[!hbt]
  \caption{
    Physical volumes $\mathcal{V(\mathrm{VMC})}$ defined in
    Eq.~(\ref{e:PhysicalVolumeV}) for the VMC geometries constructed on
    the manifolds in this study.  Also listed are the average scalar
    curvature $\langle \tilde R\,\rangle$ defined in
    Eq.~(\ref{e:AvarageRtildeDef}) for each manifold.
    \label{t:TableII} }
  \begin{center}
    \begin{tabular}{lcc}
    Manifold & $\langle \tilde R\, \rangle$
    & $\mathcal{V}(\mathrm{VMC})$ \\
    \hline
    $G2\times S1$              & -2.97 & 8.20 \\
    $L(8,3)$                   &  2.66 & 8.23 \\
    $S2\times S1$              &  2.69 & 9.39 \\
    $SFS[S2:(2,1)(2,1)(2,-1)]$ &  2.66 & 8.23 \\
  \end{tabular}
  \end{center}
\end{table}

\begin{figure}[!h]
  \centering
    \includegraphics[width=0.44\textwidth]{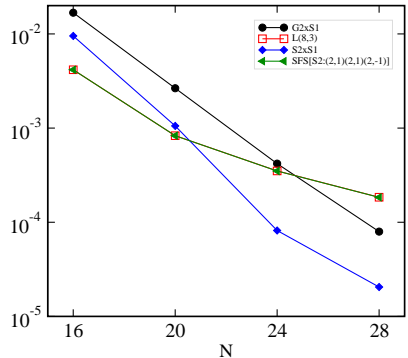}
  \caption{\label{f:VMCEinsteinConstraints}Norm of the Einstein
    constraints, $\mathcal{C}$, for the numerical VMC initial data solutions.}
\end{figure}

\begin{figure}[!h]
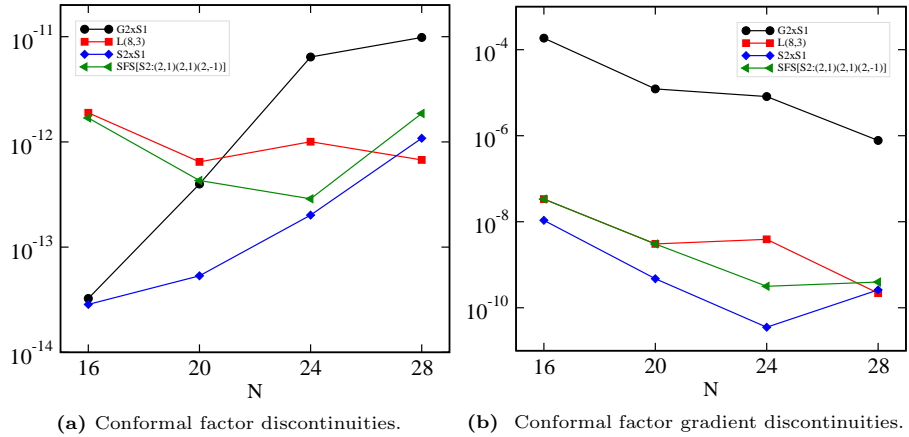

  \begin{subfigure}{0.5\textwidth} 
    \centering 
    \includegraphics[width=0.98\textwidth]{Fig4.2a.eps}
    \caption{\label{f:VMCConformalFactor}Conformal factor
      discontinuities.}
  \end{subfigure}
  \begin{subfigure}{0.5\textwidth}
    \centering
    \includegraphics[width=0.98\textwidth]{Fig4.2b.eps}
    \caption{\label{f:VMCConformalFactorGradient}
      Conformal factor gradient discontinuities.}
  \end{subfigure}
  \caption{Norms of the interface discontinuities in the conformal factor
    and its gradient as a function of spatial resolution $N$ for the
    numerical VMC solutions.
    \label{f:VMCDistontinuities}}
\end{figure}

The VMC Eqs.~(\ref{e:SimpleConstraintS2}) and
(\ref{e:SimpleConstraintV2}) are a much more complicated system than
the simple scalar CMC Eq.~(\ref{e:CMCConstraintS}).  Consequently the
numerical convergence is significantly slower.  This inefficiency made
it impractical to consider solutions with numerical resolutions larger
than $N=28$ for this study.  The $N\leq 28$ solutions took several
months running in parallel (one processor for each cubic region) to
achieve a satisfactory level of convergence.
Figure~\ref{f:VMCEinsteinConstraints} shows the norm of the Einstein
constraints Eq.~(\ref{e:ConstraintNorm}) for these numerical VMC
solutions, which are similar in size to the CMC constraint norms at
the same resolutions in Fig.~\ref{f:CMCEinsteinConstraints}.
Figure~\ref{f:VMCDistontinuities} shows the norms of the
discontinuities in the conformal factor and its gradient across the
boundary interfaces between the multicube regions.  The sizes of these
discontinuities are also comparable to those for the CMC solutions at
the same numerical resolutions in Fig.~\ref{f:CMCDiscontinuities}.

\section{Discussion}
\label{s:Discussion}

This paper outlines a basic framework for finding numerical solutions
to the Einstein constraint equations on manifolds with non-trivial
topologies.  These ideas are illustrated here using simple constant
mean curvature and variable mean curvature numerical solutions on
several different compact orientable manifolds.  The constant mean
curvature solutions found here have constant scalar curvatures and are
therefore solutions to the Yamabe problem on these manifolds as well.
The one feature of these numerical examples that was surprising (to
us) was the extreme inefficiency of our numerical elliptic solver.
Some of the numerical VMC solutions presented here required running
for months in parallel on a reasonably fast multiprocessor computer.
We plan to study ways to improve this efficiency in a future project
so that more cosmologically interesting solutions can be obtained and
studied on a larger collection of manifolds.  We plan to explore a
variety of ways this might be done, e.g. through more efficient
utilization of the PETSC solvers, by finding and implementing more
efficient numerical methods for solving elliptic equations than those
available in the SpEC code, or by finding different formulations of
the Einstein constraints that can be solved numerically more
efficiently.

\appendix
\section{Appendix: Multicube Structures}
\label{s:MulticubeStructures}
\counterwithin*{figure}{section}
\counterwithin*{table}{section}

A multicube structure consists of a set of non-overlapping cubes,
${\cal B}_A$, that cover the manifold, and a set of maps
$\Psi^{A\alpha}_{B\beta}$ that identify the faces of neighboring
cubes.  The interface boundary maps used here (written in terms of the
global Cartesian coordinates used for the multicube structure) take
points, $x^i_B$, on the interface boundary
$\partial_\beta\mathcal{B}_B$ of region $\mathcal{B}_B$ to the
corresponding points, $x^i_A$, in the boundary
$\partial_\alpha\mathcal{B}_A$ of region $\mathcal{B}_A$ in the
following way,
\begin{eqnarray}
x^i_A = c^i_A +f^i_{\alpha} + C_{B\beta\,j}^{A\alpha\, i}(x^j_B
 - c^j_B-f^j_{\beta}).
\label{e:CoordinateMap}
\end{eqnarray}
The vectors $\vec c_A+\vec f_{\alpha}$ and $\vec c_B+\vec f_{\beta}$
are the locations of the centers of the $\partial_\alpha
\mathcal{B}_A$ and $\partial_\beta \mathcal{B}_B$ faces respectively,
and ${\mathbf C}_{B\beta}^{A\alpha}$ is the combined
rotation/reflection matrix needed to orient the faces properly.

The multicube structures for two of the manifolds included in this
study, $L(8,3)$ and $SFS[S2:(2,1)(2,1)(2,-1)]$, were derived using the
methods described in \cite{Lindblom2022} from the triangulations of
these manifolds given in the Regina catalog of compact orientable
three-manifolds~\cite{Regina}.  The multicube structure for $L(8,3)$
is given here in Table~\ref{t:L83Table}.  The multicube structure for
$SFS[S2:(2,1)(2,1)(2,-1)]$ was published previoulsy in Table D.8 in
\cite{Lindblom2022}, so it is not reproduced here.  The multicube
structures for the other two manifolds included in this study,
$S2\times S1$ and $G2\times S1$ where constructed by hand.  The
multicube structure for $S2\times S1$ was published previously in
Table A.3 in \cite{Lindblom2013}.  The multicube structure used here
for $G2\times S1$ is based on the eight-region representation of the
two-manifold $G2$ in Appendix B.5 in \cite{Lindblom2015}.  The
resulting three-dimensional multicube structure is given here in
Table~\ref{t:TableG2xS1}.

The following tables include lists of the cubic regions,
$\mathcal{B}_A$, used to cover the manifold in each structure, the
vectors $\vec c_A$ that define the locations of the centers of these
regions in $\mathbb{R}^3$, and the rotation/reflection matrices
${\mathbf C}_{B\beta}^{A\alpha}$ needed to transform each cube face
into the face of its neighbor.\footnote{The vectors $\vec f_\alpha$
are the relative positions of the center of the $A\{\alpha\}$ cube
face with the center of region $\mathcal{B}_A$.  These vectors are the
same for all the cubic regions, and are given explicitly in
\cite{Lindblom2013} so they are not repeated here.}  The
identification of the $\partial_\beta \mathcal{B}_B$ face with the
$\partial_\alpha \mathcal{B}_A$ face is indicated in the tables by
$\{\alpha A\}\leftrightarrow \{\beta B\}$.  The notation $\mathbf{I}$
in these tables indicates the identity matrix, while
$\mathbf{R}_{\alpha}$ indicates the $+\pi/2$ rotation about the
outward directed normal to the $\{\alpha\}$ cube face.
\begin{table}[h]
  \tiny
  \renewcommand{\arraystretch}{1.2}
  \begin{center}
    \caption{ Multicube representation of the Regina triangulation of
      the manifold $L(8,3)$.  Multicube Structure: region center
      locations $\vec c_A$, region face identifications, $\{\alpha
      \,A\} \leftrightarrow \{\beta\, B\}$, and the rotation matrices
      for the associated interface maps, ${\bf C}_{A\alpha}^{B\beta}$.
    \label{t:L83Table}
    }
    \begin{tabular}{c|c|c|c|c}
      \toprule
      && $\alpha=-x$ & $\alpha=+x$ & $\alpha=-y$ \\
      A  & $\vec c_A$
      &$B\,\,\beta\,\,\,{\mathbf C}_{A\alpha}^{B\beta}$
      &$B\,\,\beta\,\,\,{\mathbf C}_{A\alpha}^{B\beta}$
      &$B\,\,\beta\,\,\,{\mathbf C}_{A\alpha}^{B\beta}$\\
      \midrule
      $0.0$
      & $(0, 0, 0)$
      & $0.3+z\,\,\,\mathbf{R}_{-y} \mathbf{R}_{-x}$
      & $0.1-x\,\,\,\mathbf{I}$
      & $1.2+y\,\,\,\mathbf{I}$
      \\
      $0.1$
      & $(L, 0, 0)$
      & $0.0+x\,\,\,\mathbf{I}$
      & $0.2-x\,\,\,\mathbf{R}_{+x}$
      & $1.1+x\,\,\,\mathbf{R}_{-z}$
      \\
      $0.2$
      & $(0, L, 0)$
      & $0.1+x\,\,\,\mathbf{R}_{-x}$
      & $0.1+y\,\,\,\mathbf{R}_{-z}$
      & $0.0+y\,\,\,\mathbf{I}$
      \\
      $0.3$
      & $(0, 0, L)$
      & $0.2+y\,\,\,\mathbf{R}^2_{+x} \mathbf{R}_{-z}$
      & $0.1+z\,\,\,\mathbf{R}_{+y}$
      & $1.3+z\,\,\,\mathbf{R}_{+x}$
      \\
      $1.0$
      & $(3L, 0, 0)$
      & $0.0-z\,\,\,\mathbf{R}_{+y}$
      & $1.1-x\,\,\,\mathbf{I}$
      & $1.1-z\,\,\,\mathbf{R}_{-x} \mathbf{R}_{-y}$
      \\
      $1.1$
      & $(4L, 0, 0)$
      & $1.0+x\,\,\,\mathbf{I}$
      & $0.1-y\,\,\,\mathbf{R}_{+z}$
      & $1.2-z\,\,\,\mathbf{R}^2_{+y} \mathbf{R}_{+x}$
      \\
      $1.2$
      & $(3L, L, 0)$
      & $0.2-z\,\,\,\mathbf{R}_{+y}$
      & $1.1+y\,\,\,\mathbf{R}_{-z}$
      & $1.0+y\,\,\,\mathbf{I}$
      \\
      $1.3$
      & $(3L, 0, L)$
      & $0.1-z\,\,\,\mathbf{R}_{+y}$
      & $1.1+z\,\,\,\mathbf{R}_{+y}$
      & $1.0-z\,\,\,\mathbf{R}_{-x} \mathbf{R}_{-y}$
      \\
      \bottomrule
      \toprule
      && $\alpha=+y$ & $\alpha=-z$ & $\alpha=+z$ \\
      A  & $\vec c_A$
      &$B\,\,\beta\,\,\,{\mathbf C}_{A\alpha}^{B\beta}$
      &$B\,\,\beta\,\,\,{\mathbf C}_{A\alpha}^{B\beta}$
      &$B\,\,\beta\,\,\,{\mathbf C}_{A\alpha}^{B\beta}$\\
      \midrule
      $0.0$
      & $(0, 0, 0)$
      & $0.2-y\,\,\,\mathbf{I}$
      & $1.0-x\,\,\,\mathbf{R}_{-y}$
      & $0.3-z\,\,\,\mathbf{I}$
      \\
      $0.1$
      & $(L, 0, 0)$
      & $0.2+x\,\,\,\mathbf{R}_{+z}$
      & $1.3-x\,\,\,\mathbf{R}_{-y}$
      & $0.3+x\,\,\,\mathbf{R}_{-y}$
      \\
      $0.2$
      & $(0, L, 0)$
      & $0.3-x\,\,\,\mathbf{R}^2_{+y} \mathbf{R}_{+z}$
      & $1.2-x\,\,\,\mathbf{R}_{-y}$
      & $0.3+y\,\,\,\mathbf{R}_{+x}$
      \\
      $0.3$
      & $(0, 0, L)$
      & $0.2+z\,\,\,\mathbf{R}_{-x}$
      & $0.0+z\,\,\,\mathbf{I}$
      & $0.0-x\,\,\,\mathbf{R}_{+y} \mathbf{R}_{+z}$
      \\
      $1.0$
      & $(3L, 0, 0)$
      & $1.2-y\,\,\,\mathbf{I}$
      & $1.3-y\,\,\,\mathbf{R}_{+x} \mathbf{R}_{-z}$
      & $1.3-z\,\,\,\mathbf{I}$
      \\
      $1.1$
      & $(4L, 0, 0)$
      & $1.2+x\,\,\,\mathbf{R}_{+z}$
      & $1.0-y\,\,\,\mathbf{R}_{+x} \mathbf{R}_{-z}$
      & $1.3+x\,\,\,\mathbf{R}_{-y}$
      \\
      $1.2$
      & $(3L, L, 0)$
      & $0.0-y\,\,\,\mathbf{I}$
      & $1.1-y\,\,\,\mathbf{R}^2_{+z} \mathbf{R}_{-x}$
      & $1.3+y\,\,\,\mathbf{R}_{+x}$
      \\
      $1.3$
      & $(3L, 0, L)$
      & $1.2+z\,\,\,\mathbf{R}_{-x}$
      & $1.0+z\,\,\,\mathbf{I}$
      & $0.3-y\,\,\,\mathbf{R}_{-x}$
      \\
      \bottomrule
    \end{tabular}
  \end{center}
\end{table}

\begin{table}[htb]
\scriptsize
  \renewcommand{\arraystretch}{1.5}
\begin{center}
  \caption{ Multicube representation of the product space G2$\times$S1
    constructed from the genus number $N_g=2$ two-dimensional compact
    orientable manifold. Multicube Structure: region center locations
    $\vec c_A$, region face identifications, $\{\alpha \,A\}
    \leftrightarrow \{\beta\, B\}$, and the rotation matrices for the
    associated interface maps, ${\bf C}_{A\alpha}^{B\beta}$.
    \label{t:TableG2xS1}
  }
\begin{tabular}{c|c|c|c|c|c|c|c}
  \toprule
  && $\alpha=-x$ & $\alpha=+x$ & $\alpha=-y$ & $\alpha=+y$ & $\alpha=-z$ &
  $\alpha=+z$ \\
A  & $\vec c_A$
&$B\,\,\beta\,\,\,{\mathbf C}_{A\alpha}^{B\beta}$
&$B\,\,\beta\,\,\,{\mathbf C}_{A\alpha}^{B\beta}$
&$B\,\,\beta\,\,\,{\mathbf C}_{A\alpha}^{B\beta}$
&$B\,\,\beta\,\,\,{\mathbf C}_{A\alpha}^{B\beta}$
&$B\,\,\beta\,\,\,{\mathbf C}_{A\alpha}^{B\beta}$
&$B\,\,\beta\,\,\,{\mathbf C}_{A\alpha}^{B\beta}$\\
\midrule
$1$
& $(L,2L,0)$
& $8+x\,\,\,\mathbf{I}$ & $8-x\,\,\,\mathbf{I}$
& $2+y\,\,\,\mathbf{I}$ & $4-y\,\,\,\mathbf{I}$
& $1+z\,\,\,\mathbf{I}$ & $1-z\,\,\,\mathbf{I}$
\\
$2$
& $(L,L,0)$
& $7+x\,\,\,\mathbf{I}$ & $4+x\,\,\,\mathbf{R}^2_{+z}$
& $3+y\,\,\,\mathbf{I}$ & $1-y\,\,\,\mathbf{I}$
& $2+z\,\,\,\mathbf{I}$ & $2-z\,\,\,\mathbf{I}$
\\
$3$
& $(L,0,0)$
& $6+x\,\,\,\mathbf{I}$ & $6-x\,\,\,\mathbf{I}$
& $4+y\,\,\,\mathbf{I}$ & $2-y\,\,\,\mathbf{I}$
& $3+z\,\,\,\mathbf{I}$ & $3-z\,\,\,\mathbf{I}$
\\
$4$
& $(L,-L,0)$
& $5+x\,\,\,\mathbf{I}$ & $2+x\,\,\,\mathbf{R}^2_{-z}$
& $1+y\,\,\,\mathbf{I}$ & $3-y\,\,\,\mathbf{I}$
& $4+z\,\,\,\mathbf{I}$ & $4-z\,\,\,\mathbf{I}$
\\
$5$
& $(0,-L,0)$
& $7-x\,\,\,\mathbf{R}^2_{+z}$ & $4-x\,\,\,\mathbf{I}$
& $8+y\,\,\,\mathbf{I}$ & $6-y\,\,\,\mathbf{I}$
& $5+z\,\,\,\mathbf{I}$ & $5-z\,\,\,\mathbf{I}$
\\
$6$
& $( 0, 0, 0)$
& $3+x\,\,\,\mathbf{I}$ & $3-x\,\,\,\mathbf{I}$
& $5+y\,\,\,\mathbf{I}$ & $7-y\,\,\,\mathbf{I}$
& $6+z\,\,\,\mathbf{I}$ & $6-z\,\,\,\mathbf{I}$
\\
$7$
& $(0,L,0)$
& $5-x\,\,\,\mathbf{R}^2_{-z}$ & $2-x\,\,\,\mathbf{I}$
& $6+y\,\,\,\mathbf{I}$ & $8-y\,\,\,\mathbf{I}$
& $7+z\,\,\,\mathbf{I}$ & $7-z\,\,\,\mathbf{I}$
\\
$8$
& $(0,2L,0)$
& $1+x\,\,\,\mathbf{I}$ & $1-x\,\,\,\mathbf{I}$
& $7+y\,\,\,\mathbf{I}$ & $5-y\,\,\,\mathbf{I}$
& $8+z\,\,\,\mathbf{I}$ & $8-z\,\,\,\mathbf{I}$
\\
\bottomrule
 \end{tabular}
\end{center}
\end{table}

\section*{Acknowledgments}
We thank James Isenberg for comments and suggestions that helped us
improve the clarity of this article.  F. Z. was supported by the
National Natural Science Foundation of China grants 12073005,
12021003, 11503003 and 11633001, and the Interdiscipline Research
Funds of Beijing Normal University.  L. L. was supported in part by
NSF grant 2012857 to the University of California at San Diego.

\section*{Data Availability} 
The authors will attempt to honor reasonable requests for the datasets
generated and analyzed as part this study.


\providecommand{\newblock}{}

\end{document}